\documentclass[12pt,twoside]{article}

\evensidemargin=\oddsidemargin
\def\Title#1#2#3{%
    \baselineskip=18pt
    \begin{center}
          {\large\bf{#1} \\ }
          \bigskip\bigskip
          {#2} \\
          {#3} \\
    \end{center}}
\long\def\Abstract#1{%
         \bigskip
         \parbox{0.93\textwidth}{%
                 \begin{center}
                       {\bf Abstract} \\
                 \end{center}
                 \medskip{\baselineskip=14pt #1}
                 \vss}
         \bigskip}

\makeatletter
\renewcommand{\section}%
 {\@startsection{section}{1}{0pt}%
  {-3.25ex plus -1ex minus -.2ex}{1.5ex plus .2ex}%
  {\vspace*{5mm}\raggedright\large\bf }}
\renewcommand{\thesection}{\arabic{section}.}
\@addtoreset{equation}{section}
\renewcommand{\@eqnnum}{(\thesection\theequation)}
\renewcommand{\p@equation}{\thesection}
\makeatother

\begin{document}

\Title{Hamiltonian formulation for the theory of gravity\\
and canonical transformations in extended phase space}%
{T. P. Shestakova}%
{Department of Theoretical and Computational Physics,
Southern Federal University,\\
Sorge St. 5, Rostov-on-Don 344090, Russia \\
E-mail: {\tt shestakova@sfedu.ru}}

\Abstract{A starting point for the present work was the statement recently discussed in the literature that two Hamiltonian formulations for the theory of gravity, the one proposed by Dirac and the other by Arnowitt -- Deser -- Misner, may not be related by a canonical transformation. In its turn, it raises a question about the equivalence of these two Hamiltonian formulations and their equivalence to the original formulation of General Relativity. We argue that, since the transformation from components of metric tensor to the ADM variables touches gauge degrees of freedom, which are non-canonical from the point of view of Dirac, the problem cannot be resolved in the limits of the Dirac approach. The proposed solution requires the extension of phase space by treating gauge degrees of freedom on an equal footing with other variables and introducing missing velocities into the Lagrangian by means of gauge conditions in differential form. We illustrate with a simple cosmological model the features of Hamiltonian dynamics in extended phase space. Then, we give a clear proof for the full gravitational theory that the ADM-like transformation is canonical in extended phase space in a wide enough class of possible parametrizations.}

\section{Introduction} It is generally accepted that the problem of formulating Hamiltonian dynamics for systems with constraints has been solved by Dirac in his seminal papers \cite{Dirac1,Dirac2}. It was Dirac who pointed to the importance of Hamiltonian formulation for any dynamical theory before its quantization \cite{Dirac3}. Other approaches, such as the Batalin -- Fradkin -- Vilkovisky (BFV) path integral approach \cite{BFV1,BFV2,BFV3} follow the Dirac one in what concerns the rule of constructing a Hamiltonian and the role of constraints as generators of transformations in phase space. It is believed that Dirac generalized Hamiltonian dynamics is equivalent to Lagrangian dynamics of original theory. However, even for electrodynamics the constraints do not generate a correct transformation for zero component of vector potential, $A_0$. The same situation we face in General Relativity, since gravitational constraints cannot produce correct transformations for $g_{00}$, $g_{0\mu}$ components of metric tensor. In fact, it means that the group of transformations generated by constraints differs from the group of gauge transformations of the original theory. Some authors have tried to remedy this shortcoming by modifying the Dirac approach and proposing some special prescriptions how the generator should be constructed (see, for example, \cite{Cast,BRR}). Until now this problem has not attracted much attention mainly because that it touches only transformations of gauge variables which, according to conventional viewpoint, are redundant and must not affect the physical content of the theory. It will be demonstrated in this paper that the role of gauge degrees of freedom may be more significant that it is usually thought, and the difference in the groups of transformations is the first indication to the inconsistence of the theory.

Historically, while constructing Hamiltonian dynamics for gravitational field theorists used various parametrizations of gravitational variables. Dirac dealt with original variables, which are components of metric tensor \cite{Dirac3}, whereas the most famous parametrization is probably that of Arnowitt -- Deser -- Misner (ADM) \cite{ADM}, who expressed $g_{00}$, $g_{0\mu}$ through the lapse and shift functions. To give another example, let us mention the work by Faddeev \cite{Faddeev} where quite specific variables $\lambda^0=1/h^{00}+1$, $\lambda^i=h^{0i}/h^{00}$, $q^{ij}=h^{0i}h^{0j}-h^{00}h^{ij}$, $h^{\mu\nu}=\sqrt{-g}g^{\mu\nu}$ were introduced. From the point of view of Lagrangian formalism, all the parametrizations are rightful, and the correspondent formulations are equivalent. Meanwhile, it has been shown in \cite{KK} that components of metric tensor and the ADM variables are not related by a canonical transformation. In other words, it implies that the Dirac Hamiltonian formulation for gravitation and the ADM one are not equivalent, though it is believed that each of them is equivalent to the Einstein (Lagrangian) formulation. There exists the contradiction that again witnesses about the incompleteness of the theoretical foundation.

The purpose of the present paper is to demonstrate that this contradiction can be resolved if one treats gauge gravitational degrees of freedom on an equal footing with physical variables in extended phase space. The idea of extended phase space was put forward by Batalin, Fradkin and Vilkovisky \cite{BFV1,BFV2,BFV3} who included integration over gauge and ghost degrees of freedom in their definition of path integral. However, in their approach gauge variables were still considered as non-physical, secondary degrees of freedom playing just an auxiliary role in the theory. To construct Hamiltonian dynamics for a constrained system which would be completely equivalent to Lagrangian formulation, we need to take yet another step: we should introduce into the Lagrangian missing velocities corresponding to gauge variables by means of special (differential) gauge conditions. It {\it actually extends} the phase space of physical degrees of freedom.

In Section 2 a mathematical formulation of the problem will be given. We shall see that non-equivalence of Hamiltonian formulations for different parametrizations prevents from constructing a generator of transformation in phase space which would produce correct transformations for any parametrizations. These ideas will be illustrated in Section 3 for a simple model with finite number of degrees of freedom. The mentioned above algorithms \cite{Cast,BRR} work correctly only for some parametrizations. One possible point of view (advocated, in particular, in \cite{KK}) is that only these parametrizations should be allowed while all other, not related with the first ones by canonical transformations, should be prohibited, including the ADM parametrization. However, imposing any limitations on admissible parametrizations or transformations does not seem to be a true solution to the problem. In Section 4 the outline of Hamiltonian dynamics in extended phase space will be presented, and in Section 5 it will be demonstrated for the full gravitational theory that different parametrizations from a wide enough class are related by canonical transformations. In particular, it will restore a legitimate status of the ADM parametrization. We shall discuss the results and future problems in Section 6.

\section{Canonical transformations in phase space}
It is generally known that for a system without constraints Lagrangian as well as Hamiltonian equations maintain their form under transformations to a new set of generalized coordinates
\begin{equation}
\label{coord-tr}
q^a=v^a(Q),
\end{equation}
where $v^a(Q)$ are invertible functions of their arguments. It is easy to see that any transformation (\ref{coord-tr}) correspond to a canonical transformation in phase space. Indeed, consider a quadratic in velocities Lagrangian
\begin{equation}
\label{lag1}
L=\frac12\;\Gamma_{ab}(q)\dot q^a\dot q^b-U(q).
\end{equation}
After the transformation (\ref{coord-tr}) the Lagrangian (\ref{lag1}) would read
\begin{equation}
\label{lag2}
L=\frac12\;\Gamma_{cd}(Q)\frac{\partial v^c}{\partial Q^a}\frac{\partial v^d}{\partial Q^b}\dot Q^a\dot Q^b-U(Q).
\end{equation}
New momenta $\{P_a\}$ are expressed through old momenta $\{p_a\}$ by relations
\begin{equation}
\label{mom-tr}
P_a=p_b\frac{\partial v^b}{\partial Q^a}.
\end{equation}
The transformation (\ref{coord-tr}), (\ref{mom-tr}) is canonical with the generating function which depends on new coordinates and old momenta,
\begin{equation}
\label{gen-f1}
\Phi(Q,\, p)=-p_a v^a(Q).
\end{equation}
The equations
\begin{equation}
\label{can-tr}
q^a=-\frac{\partial\Phi}{\partial p_a};\qquad
P^a=-\frac{\partial\Phi}{\partial Q^a}
\end{equation}
reproduce exactly the transformation (\ref{coord-tr}), (\ref{mom-tr}). It is also easy to check that the transformation (\ref{coord-tr}), (\ref{mom-tr}) maintains the Poisson brackets
\begin{equation}
\label{PB1}
\{Q^a,\,Q^b\}=0,\qquad
\{P_a,\,P_b\}=0,\qquad
\{Q^a,\,P_b\}=\delta^a_b.
\end{equation}

For a system with constraints, gauge variables (i.e. the variables whose velocities cannot be expressed in terms of conjugate momenta) do not enter into the Lagrangian quadratically, and a general transformation like (\ref{coord-tr}) may not be canonical. An example can be found in the theory of gravity by the transformation from components of metric tensor to the ADM variables,
\begin{equation}
\label{ADM-tr}
g_{00}=\gamma_{ij}N^i N^j-N^2,\qquad
g_{0i}=\gamma_{ij}N^j,\qquad
g_{ij}=\gamma_{ij}.
\end{equation}
This transformation concerns gauge degrees of freedom which, from the viewpoint of Dirac, are not canonical variables at all. To pose the question, if the transformation (\ref{ADM-tr}) is canonical, {\it one should formally extend} the original phase space including into it gauge degrees of freedom and their momenta. In order to prove non-canonicity of (\ref{ADM-tr}) it is enough to check that some of the relations (\ref{PB1}) are broken. Using the transformation inverted to (\ref{ADM-tr}), one can see that $\{N,\,\Pi^{ij}\}\ne 0$, where $\Pi^{ij}$ are the momenta conjugate to $\gamma_{ij}$ (see Equation (152) in \cite{KK}). More generally, let us consider the ADM-like transformation
\begin{equation}
\label{ADMl-tr}
N_{\mu}=V_{\mu}(g_{0\nu},\, g_{ij}),\qquad
\gamma_{ij}=g_{ij}.
\end{equation}
Here $V_{\mu}$ are some functions of components of metric tensor (but $N_{\mu}$ ought not to form 4-vector). A feature of this transformation is that space components of metric tensor remain unchanged, and so do their conjugate momenta: $\Pi^{ij}=p^{ij}$. Then
\begin{equation}
\label{ADM-PB}
\left.\{N_{\mu},\,\Pi^{ij}\}\right|_{g_{\nu\lambda},p^{\rho\sigma}}=\frac{\partial V_{\mu}}{\partial g_{ij}}.
\end{equation}
It is equal to zero if only the functions $V_{\mu}$ do not depend on $g_{ij}$. This is quite a trivial case when old gauge variables are expressed through some new gauge variables only, and the ADM transformation (\ref{ADM-tr}) does not belong to this class.

One could pose the question: Is it worth considering the equivalence of different formulations in extended phase space? Would not it better to restrict ourself by transformations in phase space of original canonical variables in the sense of Dirac? In the second case, we can prove the equivalence of equation of motion in Lagrangian and Hamiltonian formalism, however, we have to fix a form of gravitational constraints by forbidding any reparametrizations of gauge variables. Determination of the constraints' form is of importance for a subsequent procedure of quantization which gives rise to the problem of parametrization noninvariance (see, for example, \cite{Hall}). From the viewpoint of subsequent quantization, the ADM parametrization is more preferable, since the constraints do not depend on gauge variables in this case. I would like to emphasize that there are no solid grounds for fixing the form of the constraints, and, as we shall see in this paper, extension of phase space enables us to solve the problem of equivalence of Lagrangian and Hamiltonian formalism for gravity without any restriction on parametrizations.

As it has been already mentioned, the constraints, being considered as generators of transformations in phase space, do not produce correct transformation for all gravitational variables. To ensure the full equivalence of two formulations one has to modify the Dirac prescription, according to which the generator must be a linear combination of constraints, and replace it by a more sophisticated algorithm. The known algorithms, firstly, are relied upon algebra of constraints and, secondly, require extension of phase space. Indeed, a transformation for a variable $q^a$ produced by any generator $G$ in phase space reads
\begin{equation}
\label{arb-PB}
\delta q^a=\{q^a,\,G\}.
\end{equation}
So, to generate correct transformations for gauge variables the Poisson brackets should be defined in extended phase space. Again, the dependence of the algorithm on the algebra of constraints together with non-canonicity of the transformations like (\ref{ADMl-tr}) leads to the fact that the algorithm would work only for a limited class of parametrizations. Thus, non-equivalence of Hamiltonian formulations for different parametrizations, resulting in different algebra of constraints, prevents from constructing the generator which would produce correct transformations for any parametrizations. In the next section we shall illustrate it making use of the algorithm \cite{Cast}, for a simple model with finite number of degrees of freedom.

\section{The generator of gauge transformation: a simple example}
Now we shall consider a closed isotropic cosmological model with the Lagrangian
\begin{equation}
\label{l1}
L_1=-\frac12\frac{a\dot a^2}N+\frac12 Na.
\end{equation}
This model is traditionally described in the ADM variables ($N$ is the lapse function, $a$ is the scale factor). For our purpose, it is more convenient to go to a new variable $\mu=N^2$ which corresponds to $g_{00}$. So the Lagrangian is
\begin{equation}
\label{l2}
L_2=-\frac12\frac{a\dot a^2}{\sqrt{\mu}}+\frac12\sqrt{\mu}\,a.
\end{equation}
The canonical Hamiltonian constructed according to the rule $H=p_a\dot q^a-L$, where $\{p_a,\;q^a\}$ are pairs of variables called canonical in the sense that all the velocities $\dot q^a$ can be expressed through conjugate momenta, for our model is
\begin{equation}
\label{Ham1}
H_C=p\dot a-L_2=-\frac12\frac{\sqrt{\mu}}a\;p^2-\frac12\sqrt{\mu}\,a
\end{equation}
($p$ is the momentum conjugate to the scale factor). However, some authors include into the form $p_a\dot q^a$ also gauge variables and their momenta which are non-canonical variables in the above sense. Then we have the so-called total Hamiltonian which for our model takes the form
\begin{equation}
\label{Ham2}
H_T=\pi\dot\mu+p\dot a-L_2=\pi\dot\mu-\frac12\frac{\sqrt{\mu}}a\;p^2-\frac12\sqrt{\mu}\,a
\end{equation}
($\pi$ is the momentum conjugate to the gauge variable $\mu$). Making use of the total Hamiltonian implies {\it a mixed formalism} in which the Hamiltonian is written in terms of canonical coordinates and momenta but as well of velocities that cannot be expressed through the momenta. Nevertheless, this very Hamiltonian plays the central role in the algorithm suggested in \cite{Cast} while the canonical Hamiltonian (\ref{Ham1}) will not lead to the correct result.

In \cite{Cast} the generator of gauge transformations is sought in the form
\begin{equation}
\label{gen.gen}
G=\sum\limits_n\theta_{\mu}^{(n)}G_n^{\mu},
\end{equation}
where $G_n^{\mu}$ are first class constraints, $\theta_{\mu}^{(n)}$ are the $n$th order time derivatives of the gauge parameters $\theta_{\mu}$. In the theory of gravity the variations of $g_{\mu\nu}$ involve first order derivatives of gauge parameters, thus the generator is
\begin{equation}
\label{grav.gen}
G=\theta_{\mu}G_0^{\mu}+\dot\theta_{\mu}G_1^{\mu}.
\end{equation}
$G_n^{\mu}$ satisfy the following conditions that were derived from the requirement of invariance of motion equations under transformations in phase space:
\begin{equation}
\label{gen.cond1}
G_1^{\mu}\quad{\rm are\;primary\;constraints};
\end{equation}
\begin{equation}
\label{gen.cond2}
G_0^{\mu}+\left\{G_1^{\mu},\;H\right\}\quad{\rm are\;primary\;constraints};
\end{equation}
\begin{equation}
\label{gen.cond3}
\left\{G_0^{\mu},\;H\right\}\quad{\rm are\;primary\;constraints}.
\end{equation}
In our case $\pi=0$ is the only primary constraint of the model, so that $G_1=\pi$. The secondary constraint is
\begin{equation}
\label{sec.constr1}
\dot\pi=\left\{\pi,\;H_T\right\}=-\frac{\partial H_T}{\partial\mu}
 =\frac14\frac1{a\sqrt{\mu}}\;p^2+\frac14\frac a{\sqrt{\mu}}=T.
\end{equation}
The canonical Hamiltonian (\ref{Ham1}) appears to be proportional to the secondary constraint $T$, $H_C=-2\mu T$.

The condition (\ref{gen.cond2}) becomes
\begin{equation}
\label{cond2}
G_0+\left\{\pi,\;H_T\right\}=\alpha\pi;
\end{equation}
\begin{equation}
\label{cond2_1}
G_0=-T+\alpha\pi,
\end{equation}
$\alpha$ is a coefficient that can be found from the requirement (\ref{gen.cond3}):
\begin{equation}
\label{cond3}
\left\{G_0,\;H_T\right\}=\beta\pi;
\end{equation}
\begin{eqnarray}
\left\{G_0,\;H_T\right\}
   &=&-\left\{T,\;H_T\right\}+\alpha\left\{\pi,\;H_T\right\}
    =-\left\{T,\;\pi\dot\mu-2\mu T\right\}+\alpha T\nonumber\\
\label{cond3_1}
   &=&-\left\{T,\;\pi\right\}\dot\mu+\alpha T
    =\frac1{2\mu}\;\dot\mu T+\alpha T;
\end{eqnarray}
\begin{equation}
\label{betaal}
\beta=0;\qquad
\alpha=-\frac1{2\mu}\;\dot\mu;
\end{equation}
\begin{equation}
\label{cond2_2}
G_0=-\frac1{2\mu}\;\dot\mu\pi-T.
\end{equation}
The full generator $G$ (\ref{grav.gen}) can be written as
\begin{equation}
\label{gen1}
G=\left(-\frac1{2\mu}\;\dot\mu\pi-T\right)\theta+\pi\dot\theta.
\end{equation}
The transformation of the variable $\mu$ is
\begin{equation}
\label{mu_transf}
\delta\mu=\left\{\mu,\;G\right\}
 =-\frac1{2\mu}\;\dot\mu\theta+\dot\theta.
\end{equation}
The same expression (up to the multiplier being equal to 2) can be obtained from general transformations of the metric tensor,
\begin{equation}
\label{g_transf}
\delta g_{\mu\nu}
 =\theta^{\lambda}\partial_{\lambda}g_{\mu\nu}
 +g_{\mu\lambda}\partial_{\nu}\theta^{\lambda}
 +g_{\nu\lambda}\partial_{\mu}\theta^{\lambda};
\end{equation}
\begin{equation}
\label{g00_transf}
\delta g_{00}
 =\dot g_{00}\theta^0+2g_{00}\dot\theta^0,
\end{equation}
if one keeps in mind that $g_{00}=\mu$ and in the above formulas $\theta=\theta_0=g_{00}\theta^0$.

Ir is easy to see that the correct expression (\ref{mu_transf}) is entirely due to the replacement of the canonical Hamiltonian (\ref{Ham1}) by the total Hamiltonian (\ref{Ham2}), otherwise one would miss the contribution from the Poisson bracket $\{T,\;\pi\}$ to the generator (\ref{gen1}) (see the second line of (\ref{cond3_1})).

On the other hand, making use of the total Hamiltonian may not lead to a correct result for another parametrization. Let us return to the Lagrangian (\ref{l1}). Now the total Hamiltonian is
\begin{equation}
\label{Ham3}
H_T=\pi\dot N-\frac12\frac Na\;p^2-\frac12\;N\,a
\end{equation}
Again, $\pi$ is the momentum conjugate to the gauge variable $N$, and $\pi=0$ is the only primary constraint. The secondary constraint does not depend on $N$ in this case:
\begin{equation}
\label{sec.constr2}
\dot\pi=\left\{\pi,\;H_T\right\}=-\frac{\partial H_T}{\partial N}
 =\frac1{2a}\;p^2+\frac12\; a=T,
\end{equation}
therefore, the Poisson bracket $\left\{T,\;\pi\right\}$ in (\ref{cond3_1}) is equal to zero, and one would obtain an incorrect expression for the generator,
\begin{equation}
\label{gen2}
G=-T\theta+\pi\dot\theta.
\end{equation}
It cannot produce the correct variation of $N$, that reads
\begin{equation}
\label{N_transf}
\delta N=-\dot N\theta-N\dot\theta.
\end{equation}

As we can see, this algorithm fails to produce correct results for an arbitrary parametrization. In the next section we shall construct Hamiltonian dynamics in extended phase space and discuss its features and advantages.

\section{Extended phase space: the isotropic model}
We shall consider the effective action including gauge and ghost sectors as it appears in the path integral approach to gauge field theories,
\begin{equation}
\label{full-act1}
S=\int dt \left(L_{(grav)}+L_{(gauge)}+L_{(ghost)}\right)
\end{equation}

As was mentioned above, it is not enough just to extend pase space by including formally gauge degrees of freedom in it. One should also introduce missing velocities into the Lagrangian. It can be done by means of special (differential) gauge conditions that {\it actually extends} the phase space and enables one to avoid the mixed formalism. For our model (\ref{l1}) the equation $N=f(a)$ determines in a general form a relation between the only gauge variable $N$ and the scale factor $a$. The differential form of this relation is
\begin{equation}
\label{gfix}
\dot N=\frac{df}{da}\;\dot a.
\end{equation}
The ghost sector of the model reads
\begin{equation}
\label{Lghost}
L_{(ghost)}=\dot{\bar\theta}N\dot\theta
 +\dot{\bar\theta}\left(\dot N-\frac{df}{da}\;\dot a\right)\theta,
\end{equation}
so that
\begin{eqnarray}
L&=&-\frac12\frac{a\dot a^2}N+\frac12 Na
 +\lambda\left(\dot N-\frac{df}{da}\;\dot a\right)
 +\dot{\bar\theta}\left(\dot N-\frac{df}{da}\;\dot a\right)\theta
 +\dot{\bar\theta}N\dot\theta=\nonumber\\
\label{Lagr3}
&=&-\frac12\frac{a\dot a^2}N+\frac12 Na
 +\pi\left(\dot N-\frac{df}{da}\;\dot a\right)
 +\dot{\bar\theta}N\dot\theta.
\end{eqnarray}
The conjugate momenta are:
\begin{equation}
\label{mom1}
\pi=\lambda+\dot{\bar\theta}\theta;\quad
p=-\frac{a\dot a}N-\pi\frac{df}{da};\quad
\bar{\cal P}=N\dot{\bar\theta};\quad
{\cal P}=N\dot\theta.
\end{equation}

Let us now go to a new variable
\begin{equation}
\label{ch_var1}
N=v(\tilde N,a).
\end{equation}
At the same time, the rest variables are unchanged:
\begin{equation}
\label{ch_var2}
a=\tilde a;\quad
\theta=\tilde\theta;\quad
\bar\theta=\tilde{\bar\theta}.
\end{equation}
It is the analog of the transformation from the original gravitational variables $g_{\mu\nu}$ to the ADM variables. Indeed, in the both cases only gauge variables are transformed while the rest variables remain unchanged. After the change (\ref{ch_var1}) the Lagrangian is written as (below we shall omit the tilde over $a$ and ghost variables which remain unchanged)
\begin{equation}
\label{Lagr4}
L=-\frac12\;\frac{a\dot a^2}{v(\tilde N,a)}
  +\frac12\; v(\tilde N,a)\;a
  +\pi\left(\frac{\partial v}{\partial\tilde N}\;\dot{\tilde N}
  +\frac{\partial v}{\partial a}\;\dot a-\frac{d f}{da}\;\dot a\right)
 +v(\tilde N,a)\;\dot{\bar\theta}\dot\theta.
\end{equation}
The new momenta are:
\begin{equation}
\label{mom2_1}
\tilde\pi=\pi\frac{\partial v}{\partial\tilde N};\qquad
\tilde p=-\frac{a\dot a}{v(\tilde N,a)}
  +\pi\frac{\partial v}{\partial a}-\pi\frac{d f}{da}
 =p+\pi\frac{\partial v}{\partial a};
\end{equation}
\begin{equation}
\label{mom2_2}
\tilde{\bar{\cal P}}=v(\tilde N,a)\;\dot{\bar\theta}=\bar{\cal P};\qquad
\tilde{\cal P}=v(\tilde N,a)\;\dot{\theta}={\cal P}.
\end{equation}
It is easy to demonstrate that the transformations (\ref{ch_var1}), (\ref{ch_var2}), (\ref{mom2_1}), (\ref{mom2_2}) are canonical in extended phase space. The generating function will depend on new coordinates and old momenta,
\begin{equation}
\label{gen-f2}
\Phi\left(\tilde N,\;\tilde a,\;\tilde{\bar\theta},\;\tilde\theta,\;
  \pi,\;p,\;\bar{\cal P},\;{\cal P}\right)
 =-\pi\, v(\tilde N,\tilde a)
  -p\,\tilde a-\bar{\cal P}\,\tilde\theta
  -\tilde{\bar\theta}\,{\cal P}.
\end{equation}
One can see that the generating function has the same form as in (\ref{gen-f1}). The relations
\begin{equation}
\label{tr1}
N=-\frac{\partial\Phi}{\partial\pi};\qquad
a=-\frac{\partial\Phi}{\partial p};\qquad
\tilde\pi=-\frac{\partial\Phi}{\partial\tilde N};\qquad
\tilde p=-\frac{\partial\Phi}{\partial\tilde a};
\end{equation}
\begin{equation}
\label{tr2}
\theta=-\frac{\partial\Phi}{\partial\bar{\cal P}\vphantom{\sqrt N}};\qquad
\bar\theta=-\frac{\partial\Phi}{\partial{\cal P}};\qquad
\tilde{\cal P}=-\frac{\partial\Phi}{\partial\tilde{\bar\theta}};\qquad
\tilde{\bar{\cal P}}=-\frac{\partial\Phi}{\partial\tilde\theta}
\end{equation}
give exactly the transformation (\ref{ch_var1}), (\ref{ch_var2}), (\ref{mom2_1}), (\ref{mom2_2}). On the other hand, one can check that Poisson brackets among all phase variables maintain their canonical form.

Now we are going to write down equations of motion in extended phase space. Firstly, we rewrite the Lagrangian (\ref{Lagr4}) through the momentum $\tilde\pi$.
\begin{eqnarray}
L&=&-\frac12\;\frac{a\dot a^2}{v(\tilde N,a)}+\frac12\; v(\tilde N,a)\; a\nonumber\\
\label{Lagr5}
 &+&\tilde\pi\left[\dot{\tilde N}
  +\left(\frac{\partial v}{\partial\tilde N}\right)^{-1}\frac{\partial v}{\partial a}\;\dot a
  -\left(\frac{\partial v}{\partial\tilde N}\right)^{-1}\frac{d f}{da}\;\dot a\right]
 +v(\tilde N,a)\;\dot{\bar\theta}\dot\theta.
\end{eqnarray}

The variation of (\ref{Lagr5}) gives, accordingly, the equation of motion (\ref{Le1}), the constraint (\ref{Le2}), the gauge condition (\ref{Le3}) and the ghost equations (\ref{Le4}) -- (\ref{Le5}):
\begin{eqnarray}
\frac{a\ddot a}{v(\tilde N,a)}
 &+&\frac12\;\frac{\dot a^2}{v(\tilde N,a)}
  -\frac12\;\frac{a\dot a^2}{v^2(\tilde N,a)}\;\frac{\partial v}{\partial a}
  -\frac{a\dot a}{v^2(\tilde N,a)}\;\frac{\partial v}{\partial\tilde N}\dot{\tilde N}\nonumber\\
 &+&\frac12\;\frac{\partial v}{\partial a}\;a+\frac12v(\tilde N,a)
  -\dot{\tilde\pi}\left(\frac{\partial v}{\partial\tilde N}\right)^{-1}\frac{\partial v}{\partial a}
  +\dot{\tilde\pi}\left(\frac{\partial v}{\partial\tilde N}\right)^{-1}\frac{d f}{da}\nonumber\\
 &+&\tilde\pi\left(\frac{\partial v}{\partial\tilde N}\right)^{-2}\frac{\partial^2v}{\partial\tilde N^2}\;
    \frac{\partial v}{\partial a}\;\dot{\tilde N}
  -\tilde\pi\left(\frac{\partial v}{\partial\tilde N}\right)^{-1}
    \frac{\partial^2v}{\partial\tilde N\partial a}\;\dot{\tilde N}\nonumber\\
\label{Le1}
 &-&\tilde\pi\left(\frac{\partial v}{\partial\tilde N}\right)^{-2}
    \frac{\partial^2v}{\partial\tilde N^2}\;\frac{d f}{da}\;\dot{\tilde N}
  +\frac{\partial v}{\partial a}\;\dot{\bar\theta}\dot\theta=0;
\end{eqnarray}
\begin{eqnarray}
\label{Le2}
\frac12\;\frac{a\dot a^2}{v^2(\tilde N,a)}\;\frac{\partial v}{\partial\tilde N}
 &+&\frac12\;\frac{\partial v}{\partial\tilde N}\;a-\dot{\tilde\pi}
   -\tilde\pi\left(\frac{\partial v}{\partial\tilde N}\right)^{-2}\frac{\partial^2v}{\partial\tilde N^2}\;
    \frac{\partial v}{\partial a}\;\dot a
   +\tilde\pi\left(\frac{\partial v}{\partial\tilde N}\right)^{-1}
    \frac{\partial^2v}{\partial\tilde N\partial a}\;\dot a\nonumber\\
 &+&\tilde\pi\left(\frac{\partial v}{\partial\tilde N}\right)^{-2}\frac{\partial^2v}{\partial\tilde N^2}\;
    \frac{d f}{da}\;\dot a
  +\frac{\partial v}{\partial\tilde N}\;\dot{\bar\theta}\dot\theta=0;
\end{eqnarray}
\begin{equation}
\label{Le3}
\frac{\partial v}{\partial\tilde N}\;\dot{\tilde N}
  +\frac{\partial v}{\partial a}\;\dot a-\frac{d f}{da}\;\dot a=0;
\end{equation}
\begin{equation}
\label{Le4}
v(\tilde N,\;a)\;\ddot\theta
  +\frac{\partial v}{\partial\tilde N}\;\dot{\tilde N}\dot\theta
  +\frac{\partial v}{\partial a}\;\dot a\dot\theta=0;
\end{equation}
\begin{equation}
\label{Le5}
v(\tilde N,\;a)\;\ddot{\bar\theta}
  +\frac{\partial v}{\partial\tilde N}\;\dot{\tilde N}\dot{\bar\theta}
  +\frac{\partial v}{\partial a}\;\dot a\dot{\bar\theta}=0.
\end{equation}

The Hamiltonian in extended phase space looks like
\begin{eqnarray}
H&=&-\frac12\;\frac{v(\tilde N,a)}a\left[\tilde p^2
    +2\tilde p\tilde\pi\left(\frac{\partial v}{\partial\tilde N}\right)^{-1}\frac{d f}{da}
    +\tilde\pi^2\left(\frac{\partial v}{\partial\tilde N}\right)^{-2}
     \left(\frac{d f}{da}\right)^2\right.\nonumber\\
 &-&\left.2\tilde p\tilde\pi\left(\frac{\partial v}{\partial\tilde N}\right)^{-1}\frac{\partial v}{\partial a}
    -2\tilde\pi^2\left(\frac{\partial v}{\partial\tilde N}\right)^{-2}
     \frac{\partial v}{\partial a}\;\frac{df}{da}
    +\tilde\pi^2\left(\frac{\partial v}{\partial\tilde N}\right)^{-2}
     \left(\frac{\partial v}{\partial a}\right)^2\right]\nonumber\\
\label{Ham.EPS}
  &-&\frac12\; v(\tilde N,a)\;a+\frac1{v(\tilde N,a)}\;\bar{\cal P}{\cal P}.
\end{eqnarray}

The Hamiltonian equations in extended phase space are:
\begin{eqnarray}
\dot{\tilde p}
 &=&\frac12\left[\frac1a\frac{\partial v}{\partial a}-\frac{v(\tilde N,a)}{a^2}\right]
    \left[\tilde p+\tilde\pi\left(\frac{\partial v}{\partial\tilde N}\right)^{-1}\frac{d f}{da}
    -\tilde\pi\left(\frac{\partial v}{\partial\tilde N}\right)^{-1}\frac{\partial v}{\partial a}\right]^2\nonumber\\
 &-&\frac{v(\tilde N,a)}a\left[\tilde\pi\left(\frac{\partial v}{\partial\tilde N}\right)^{-2}
      \frac{\partial^2v}{\partial\tilde N\partial a}\;\frac{d f}{da}
     -\tilde\pi\left(\frac{\partial v}{\partial\tilde N}\right)^{-1}\frac{d^2f}{da^2}\right.\nonumber\\
 &-&\left.\tilde\pi\left(\frac{\partial v}{\partial\tilde N}\right)^{-2}
      \frac{\partial^2v}{\partial\tilde N\partial a}\;\frac{\partial v}{\partial a}
     +\tilde\pi\left(\frac{\partial v}{\partial\tilde N}\right)^{-1}\frac{\partial^2v}{\partial a^2}\right]\nonumber\\
 &\times &\left[\tilde p+\tilde\pi\left(\frac{\partial v}{\partial\tilde N}\right)^{-1}\frac{d f}{da}
      -\tilde\pi\left(\frac{\partial v}{\partial\tilde N}\right)^{-1}\frac{\partial v}{\partial a}\right]\nonumber\\
\label{He1}
 &+&\frac12\;\frac{\partial v}{\partial a}\; a+\frac12\;v(\tilde N,a)
     +\frac1{v^2(\tilde N,a)}\;\bar{\cal P}{\cal P};\\
\label{He2}
\dot a
 &=&-\frac{v(\tilde N,a)}a\left[\tilde p
    +\tilde\pi\left(\frac{\partial v}{\partial\tilde N}\right)^{-1}\frac{d f}{da}
    -\tilde\pi\left(\frac{\partial v}{\partial\tilde N}\right)^{-1}\frac{\partial v}{\partial a}\right];\\
\dot{\tilde\pi}
 &=&\frac1{2a}\;\frac{\partial v}{\partial\tilde N}\left[\tilde p
    +\tilde\pi\left(\frac{\partial v}{\partial\tilde N}\right)^{-1}\frac{d f}{da}
    -\tilde\pi\left(\frac{\partial v}{\partial\tilde N}\right)^{-1}\frac{\partial v}{\partial a}\right]^2\nonumber\\
 &-&\frac{v(\tilde N,a)}a\left[\tilde\pi\left(\frac{\partial v}{\partial\tilde N}\right)^{-2}
      \frac{\partial^2v}{\partial\tilde N^2}\;\frac{d f}{da}
     -\tilde\pi\left(\frac{\partial v}{\partial\tilde N}\right)^{-2}
      \frac{\partial^2v}{\partial\tilde N^2}\;\frac{\partial v}{\partial a}\right.\nonumber\\
 &+&\left.\tilde\pi\left(\frac{\partial v}{\partial\tilde N}\right)^{-1}
      \frac{\partial^2v}{\partial\tilde N\partial a}\right]
   \left[\tilde p+\tilde\pi\left(\frac{\partial v}{\partial\tilde N}\right)^{-1}\frac{d f}{da}
    -\tilde\pi\left(\frac{\partial v}{\partial\tilde N}\right)^{-1}\frac{\partial v}{\partial a}\right]\nonumber\\
\label{He3}
 &+&\frac12\;\frac{\partial v}{\partial\tilde N}\; a
  +\frac1{v^2(\tilde N,a)}\;\frac{\partial v}{\partial\tilde N}\;\bar{\cal P}{\cal P};\\
\dot{\tilde N}
 &=&-\frac{v(\tilde N,a)}a\left[\left(\frac{\partial v}{\partial\tilde N}\right)^{-1}\frac{d f}{da}
   -\left(\frac{\partial v}{\partial\tilde N}\right)^{-1}\frac{\partial v}{\partial a}\right]\nonumber\\
\label{He4}
 &\times &\left[\tilde p+\tilde\pi\left(\frac{\partial v}{\partial\tilde N}\right)^{-1}\frac{d f}{da}
    -\tilde\pi\left(\frac{\partial v}{\partial\tilde N}\right)^{-1}\frac{\partial v}{\partial a}\right];\\
\label{He5}
\dot{\bar{\cal P}}&=&0;\\
\label{He6}
\dot\theta &=&\frac1{v(\tilde N,a)}\;{\cal P};\\
\label{He7}
\dot{\cal P}&=&0;\\
\label{He8}
\dot{\bar\theta}&=&\frac1{v(\tilde N,a)}\;\bar{\cal P}.
\end{eqnarray}

One can check that the Hamiltonian equations (\ref{He1}) -- (\ref{He8}) are completely equivalent to the Lagrangian equations (\ref{Le1}) -- (\ref{Le5}), the constraint (\ref{He3}) and the gauge condition (\ref{He4}) being true Hamiltonian equations.

The Hamiltonian equations (\ref{He1}) -- (\ref{He8}) in extended phase space, as well as the equations (\ref{Le1}) -- (\ref{Le5}), include gauge-dependent terms. In this connection one can object that the equations are not equivalent to the original Einstein equation, which are known to be gauge-invariant. However, we remember that any solution to the gauge-invariant Einstein equation is determined up to arbitrary functions which have to be fix by a choice of a reference frame (a state of the observer). It is usually done on the final stage of solving the Einstein equations. It is important that {\it one cannot avoid fixing a reference frame to obtain a final form of the solution}. By varying the gauged action (\ref{full-act1}), in fact, we deal with a generalized mathematical problem, its generalization has come from the development of quantum field theory.

In the case of the extended set of equations (\ref{He1}) -- (\ref{He8}) (or, (\ref{Le1}) -- (\ref{Le5})) one can keep the function $f(a)$ non-fixed up to the final stage of their resolution. Further, under the conditions $\bar\pi=0$, $\theta=0$, $\bar\theta=0$ all gauge-dependent terms are excluded, and the extended set of equations is reduced to gauge-invariant equations, therefore, any solution of the Einstein equations can be found among solutions of the extended set. Solutions with non-trivial $\bar\pi$, $\theta$, $\bar\theta$ should be considered and physically interpreted separately.

One can also reveal that there exists a quantity conserved if the Hamiltonian (or, equivalently, Lagrangian) equations hold. It plays the role of the BRST generator for our model:
\begin{equation}
\label{BRST_mod}
\Omega=-H\theta-\left(\frac{\partial v}{\partial\tilde N}\right)^{-1}\tilde\pi{\cal P}.
\end{equation}
It generates correct transformations for the variables $a$, $\theta$, $\bar\theta$ and for any gauge variable $\tilde N$ given by the relation (\ref{ch_var1}),
\begin{equation}
\label{tilN_trasf}
\delta\tilde N
 =-\frac{\partial H}{\partial\tilde\pi}\;\theta
  -\left(\frac{\partial v}{\partial\tilde N}\right)^{-1}{\cal P}
 =-\dot{\tilde N}\theta
  -\left(\frac{\partial v}{\partial\tilde N}\right)^{-1}v(\tilde N,a)\;\dot\theta.
\end{equation}
In particular, for the original variable $N$ one gets the transformation (\ref{N_transf}).

\section{The canonicity of transformations in extended phase space\\
for the full gravitational theory}
In this section we shall demonstrate for the full gravitational theory that different parametrizations from a wide enough class (\ref{ADMl-tr}) are related by canonical transformations. Again, we shall start from the gauged action
\begin{equation}
\label{full-act}
S=\int d^4 x\left({\cal L}_{(grav)}+{\cal L}_{(gauge)}+{\cal L}_{(ghost)}\right)
\end{equation}

We shall use a gauge condition in a general form, $f^{\mu }(g_{\nu\lambda})=0$. The differential form of this gauge condition introduces the missing velocities and actually extends phase space,
\begin{equation}
\label{diff-g}
\frac{d}{dt}f^{\mu}(g_{\nu\lambda})=0,\qquad
\frac{\partial f^{\mu}}{\partial g_{00}}\dot g_{00}
 +2\frac{\partial f^{\mu}}{\partial g_{0i}}\dot g_{0i}
 +\frac{\partial f^{\mu}}{\partial g_{ij}}\dot g_{ij}=0.
\end{equation}
Then,
\begin{equation}
\label{L_gauge}
{\cal L}_{(gauge)}=\lambda_{\mu}\left(\frac{\partial f^{\mu}}{\partial g_{00}}\dot g_{00}
  +2\frac{\partial f^{\mu}}{\partial g_{0i}}\dot g_{0i}
  +\frac{\partial f^{\mu}}{\partial g_{ij}}\dot g_{ij}\right).
\end{equation}
Taking into account the gauge transformations,
\begin{equation}
\label{g-tr}
\delta g_{\mu\nu}=\partial_{\lambda}g_{\mu\nu}\theta^{\lambda}
  +g_{\mu\lambda}\partial_{\nu}\theta^{\lambda}
  +g_{\nu\lambda}\partial_{\mu}\theta^{\lambda},
\end{equation}
one can write the ghost sector:
\begin{equation}
\label{L_ghost}
{\cal L}_{(ghost)}=\bar\theta_{\mu}\frac{d}{dt}
  \left[\frac{\partial f^{\mu}}{\partial g_{\nu\lambda}}
   \left(\partial_{\rho}g_{\nu\lambda}\theta^{\rho}
     +g_{\lambda\rho}\partial_{\nu}\theta^{\rho}
     +g_{\nu\rho}\partial_{\lambda}\theta^{\rho}\right)\right].
\end{equation}

It is convenient to write down the action (\ref{full-act}), (\ref{L_gauge}), (\ref{L_ghost}) in the form
\begin{eqnarray}
S&=&\int d^4 x\left[{\cal L}_{(grav)}+\Lambda_{\mu}
  \left(\frac{\partial f^{\mu}}{\partial g_{00}}\dot g_{00}
   +2\frac{\partial f^{\mu}}{\partial g_{0i}}\dot g_{0i}
   +\frac{\partial f^{\mu}}{\partial g_{ij}}\dot g_{ij}\right)\right.\nonumber\\
 &-&\dot{\bar{\theta_{\mu}}}\left(\frac{\partial f^{\mu}}{\partial g_{00}}
    \left(\partial_i g_{00}\theta^i+2g_{0\nu}\dot\theta^{\nu}\right)
   +2\frac{\partial f^{\mu}}{\partial g_{0i}}
    \left(\partial_j g_{0i}\theta^{j}+g_{0\nu}\partial_i\theta^{\nu}+g_{i\nu}\dot\theta^{\nu}\right)
    \right.\nonumber\\
\label{action1}
 &+&\left.\left.\frac{\partial f^{\mu}}{\partial g_{ij}}\left(\partial_k g_{ij}\theta^k
   +g_{i\nu}\partial_j\theta^{\nu}+g_{j\nu}\partial_i\theta^{\nu}\right)\right)\right].
\end{eqnarray}
Here $\Lambda_{\mu}=\lambda_{\mu}-\dot{\bar{\theta_{\mu}}}\theta^0$. One can see that the generalized velocities enter into the bracket multiplied by $\Lambda_{\mu}$, in addition to the gravitational part ${\cal L}_{(grav)}$. This very circumstance will ensure the canonicity of the transformation to new variables.

Our goal now is to introduce new variables by
\begin{equation}
\label{new-var}
g_{0\mu}=v_{\mu}\left(N_{\nu},g_{ij}\right).\qquad
g_{ij}=\gamma_{ij};\qquad
\theta^{\mu}=\tilde\theta^{\mu};\qquad
\bar\theta_{\mu}=\tilde{\bar\theta_{\mu}}.
\end{equation}
This is the inverse transformation for (\ref{ADMl-tr}) and concerns only $g_{0\mu}$ metric components. After the transformation the action will read
\begin{eqnarray}
S&=&\int d^4 x\left[{\cal L'}_{(grav)}
   +\Lambda_{\mu}\left(\frac{\partial f^{\mu }}{\partial g_{00}}\;
     \frac{\partial v_0}{\partial N_{\nu}}\;\dot N_{\nu}
    +\frac{\partial f^{\mu}}{\partial g_{00}}\;
     \frac{\partial v_0}{\partial g_{ij}}\;\dot g_{ij}\right.\right.\nonumber\\
 &+&\left.2\;\frac{\partial f^{\mu}}{\partial g_{0i}}\;
     \frac{\partial v_i}{\partial N_{\nu}}\;\dot N_{\nu}
    +2\;\frac{\partial f^{\mu}}{\partial g_{0k}}\;
     \frac{\partial v_k}{\partial g_{ij}}\;\dot g_{ij}
    +\frac{\partial f^{\mu}}{\partial g_{ij}}\;\dot g_{ij}\right)\nonumber\\
 &-&\dot{\bar{\theta_{\mu}}}\left(\frac{\partial f^{\mu}}{\partial g_{00}}\;
     \frac{\partial v_0}{\partial N_{\nu}}\;\partial_i N_{\nu}\theta^i
    +\frac{\partial f^{\mu}}{\partial g_{00}}\;
     \frac{\partial v_0}{\partial g_{ij}}\;\partial_k g_{ij}\theta^k
    +2\;\frac{\partial f^{\mu}}{\partial g_{00}}\;v_{\nu}(N_{\lambda},g_{ij})\;
     \dot{\theta}^{\nu}\right.\nonumber\\
 &+&2\;\frac{\partial f^{\mu}}{\partial g_{0i}}\;
     \frac{\partial v_i}{\partial N_{\nu}}\;\partial_j N_{\nu}\theta^j
    +2\;\frac{\partial f^{\mu}}{\partial g_{0i}}\;
     \frac{\partial v_i}{\partial g_{jk}}\;\partial_l g_{jk}\theta^l\nonumber\\
 &+&2\;\frac{\partial f^{\mu}}{\partial g_{0i}}\left[v_{\nu}(N_{\lambda},g_{jk})\;\partial_i\theta^{\nu}
    +v_i(N_{\lambda},g_{jk})\;\dot\theta^0+g_{ij}\dot\theta^j\right]\nonumber\\
 &+&\frac{\partial f^{\mu}}{\partial g_{ij}}\left[\partial_k g_{ij}\theta^k
    +v_i(N_{\lambda},g_{kl})\;\partial_j\theta^0+g_{ik}\partial_j\theta^k\right.\nonumber\\
\label{action2}
 &+&\left.\left.\left.v_j(N_{\lambda},g_{kl})\;\partial_i\theta^0
    +g_{jk}\partial_i\theta^k\right]\right)\right]
\end{eqnarray}

We can write down the ``old'' momenta,
\begin{eqnarray}
\label{old-mom1}
\pi^{ij}&=&\frac{\partial{\cal L}_{(grav)}}{\partial\dot g_{ij}}
 +\Lambda_{\mu}\;\frac{\partial f^{\mu}}{\partial g_{ij}};\\
\label{old-mom2}
\pi^0&=&\frac{\partial{\cal L}_{(grav)}}{\partial\dot g_{00}}
 +\Lambda_{\mu}\;\frac{\partial f^{\mu}}{\partial g_{00}};\\
\label{old-mom3}
\pi^i&=&\frac{\partial{\cal L}_{(grav)}}{\partial\dot g_{0i}}
 +2\Lambda_{\mu}\;\frac{\partial f^{\mu}}{\partial g_{0i}},
\end{eqnarray}
and the ``new'' momenta are:
\begin{eqnarray}
\label{new-mom1}
\Pi^{ij}&=&\frac{\partial{\cal L'}_{(grav)}}{\partial\dot g_{ij}}
 +\Lambda_{\mu}\left(\frac{\partial f^{\mu}}{\partial g_{00}}\;
   \frac{\partial v_0}{\partial g_{ij}}
 +2\;\frac{\partial f^{\mu}}{\partial g_{0k}}\;
   \frac{\partial v_k}{\partial g_{ij}}
 +\frac{\partial f^{\mu}}{\partial g_{ij}}\right);\\
\label{new-mom2}
\Pi^0&=&\frac{\partial{\cal L'}_{(grav)}}{\partial\dot N_0}
 +\Lambda_{\mu}\left(\frac{\partial f^{\mu}}{\partial g_{00}}\;
   \frac{\partial v_0}{\partial N_0}
 +2\;\frac{\partial f^{\mu}}{\partial g_{0i}}\;
   \frac{\partial v_i}{\partial N_0}\right);\\
\label{new-mom3}
\Pi^i&=&\frac{\partial{\cal L'}_{(grav)}}{\partial\dot N_i}
 +\Lambda_{\mu}\left(\frac{\partial f^{\mu}}{\partial g_{00}}\;
  \frac{\partial v_0}{\partial N_i}
 +2\;\frac{\partial f^{\mu}}{\partial g_{0j}}\;
   \frac{\partial v_j}{\partial N_i}\right).
\end{eqnarray}

The relations between the ``old'' and ``new'' momenta:
\begin{eqnarray}
\label{relat1_1}
\Pi^{ij}&=&\pi^{ij}+\left(\pi^{\mu}
 -\frac{\partial{\cal L}_{(grav)}}{\partial\dot g_{0\mu}}\right)
   \frac{\partial v_{\mu}}{\partial g_{ij}};\\
\label{relat1_2}
\Pi^{\mu}&=&\frac{\partial{\cal L'}_{(grav)}}{\partial\dot N_{\mu}}
 +\left(\pi^{\nu}-\frac{\partial{\cal L}_{(grav)}}{\partial\dot g_{0\nu}}\right)
   \frac{\partial v_{\nu}}{\partial N_{\mu}}.
\end{eqnarray}
It is easy to check that the momenta conjugate to ghosts remain unchanged, $\tilde{\cal P}^{\mu}={\cal P}^{\mu}$,
$\tilde{\bar{\cal P}}_{\mu}=\bar{\cal P}_{\mu}$.

As any Lagrangian is determined up to total derivatives, the gravitational Lagrangian density ${\cal L}_{(grav)}$ can be modified in such a way for the primary constraints to take the form $\pi^{\mu}=0$, where $\pi^{\mu}$ are the momenta conjugate to gauge variables $g_{0\mu}$. This change of the Lagrangian density does not affect the equation of  motion. It was made by Dirac \cite{Dirac3} to simplify the calculations. A similar change of the Lagrangian density by omitting a divergence and a total time derivative was made also in the ADM paper \cite{ADM}. Therefore, one can put
\begin{equation}
\label{deriv}
\frac{\partial{\cal L}_{(grav)}}{\partial\dot g_{0\mu}}=0,\qquad
\frac{\partial{\cal L'}_{(grav)}}{\partial\dot N_{\mu}}=0.
\end{equation}
Then, the relations (\ref{relat1_1}) -- (\ref{relat1_2}) would become simpler and take the form
\begin{equation}
\label{relat2}
\Pi^{ij}=\pi^{ij}+\pi^{\mu}\frac{\partial v_{\mu}}{\partial g_{ij}};\qquad
\Pi^{\mu}=\pi^{\nu}\frac{\partial v_{\nu}}{\partial N_{\mu}}.
\end{equation}

It is easy to demonstrate that the transformations (\ref{ADMl-tr}), (\ref{relat2}) are canonical in extended phase space. The generating function again depends on new coordinates and old momenta and has the same form as for a non-constrained system (see (\ref{gen-f1}), compare also with (\ref{gen-f2})),
\begin{equation}
\label{gen-f3}
\Phi\left(N_{\mu},\;\gamma_{ij},\;\tilde\theta^{\mu},\;\tilde{\bar\theta}_{\mu},\;
   \pi^{\mu},\;\pi^{ij},\;\bar{\cal P}_{\mu},\;{\cal P}^{\mu}\right)
 =-\pi^{\mu}v_{\mu}(N_{\nu},\gamma_{ij})-\pi^{ij}\gamma_{ij}
  -\bar{\cal P}_{\mu}\tilde\theta^{\mu}-\tilde{\bar\theta}_{\mu}{\cal P}^{\mu }.
\end{equation}

Then the following relations take place
\begin{equation}
\label{can-rel1}
g_{0\mu}=-\frac{\partial\Phi}{\partial\pi^{\mu}};\qquad
g_{ij}=-\frac{\partial\Phi}{\partial\pi^{ij}};\qquad
\theta^{\mu}=-\frac{\partial\Phi}{\partial\bar{\cal P}\vphantom{\sqrt N}_{\mu}};\qquad
\bar{\theta}_{\mu}=-\frac{\partial\Phi}{\partial{\cal P}^{\mu }};
\end{equation}
\begin{equation}
\label{can-rel2}
\Pi^{\mu}=-\frac{\partial\Phi}{\partial N_{\mu}};\qquad
\Pi^{ij}=-\frac{\partial\Phi}{\partial\gamma_{ij}};\qquad
\tilde{\bar{\cal P}}_{\mu}=-\frac{\partial\Phi}{\partial\tilde\theta^{\mu}};\qquad
\tilde{\cal P}^{\mu}=-\frac{\partial\Phi}{\partial\tilde{\bar\theta}\vphantom{\sqrt N}_{\mu}},
\end{equation}
that give exactly the transformations
\begin{equation}
\label{c-rel1}
g_{0\mu}=v_{\mu}(N_{\nu},\gamma_{ij});\qquad
g_{ij}=\gamma_{ij};\qquad\qquad\qquad\qquad
\theta^{\mu}=\tilde\theta^{\mu};\qquad
\bar{\theta}_{\mu}=\tilde{\bar\theta}_{\mu };
\end{equation}
\begin{equation}
\label{c-rel2}
\Pi^{\mu }=\pi^{\nu}\frac{\partial v_{\nu}}{\partial N_{\mu}};\qquad\qquad
\Pi^{ij}=\pi^{ij}+\pi^{\mu}\frac{\partial v_{\mu}}{\partial g_{ij}};\qquad\quad
\tilde{\bar{\cal P}}_{\mu}=\bar{\cal P}_{\mu};\qquad
\tilde{\cal P}^{\mu}={\cal P}^{\mu}.
\end{equation}

We can now check if the Poisson brackets maintain their form. Differentiating the first relation in (\ref{ADMl-tr}) with respect to $g_{ij}$ one gets
\begin{equation}
\label{im-diff1}
\frac{\partial V_{\mu}}{\partial g_{ij}}
 +\frac{\partial V_{\mu}}{\partial g_{0\lambda}}\frac{\partial v_{\lambda}}{\partial g_{ij}}=0,
\end{equation}
Similarly, differentiating the same relation with respect to $N_{\nu}$ gives
\begin{equation}
\label{im-diff2}
\delta_{\mu}^{\nu}
 -\frac{\partial V_{\mu}}{\partial g_{0\lambda}}\frac{\partial v_{\lambda}}{\partial N_{\nu}}=0.
\end{equation}
Making use of (\ref{im-diff1}), (\ref{im-diff2}), it is not difficult to calculate the Poisson brackets. So, we can recalculate the bracket (\ref{ADM-PB}) to see that it will be zero in our extended phase space formalism,
\begin{eqnarray}
\left.\{N_{\mu},\,\Pi^{ij}\}\right|_{g_{\nu\lambda},p^{\rho\sigma}}
 &=&\frac{\partial N_{\mu}}{\partial g_{0\rho}}\;\frac{\partial \Pi^{ij}}{\partial\pi^{\rho}}
  +\frac{\partial N_{\mu}}{\partial g_{kl}}\;\frac{\partial\Pi^{ij}}{\partial\pi^{kl}}
  =\left\{V_{\mu}(g_{0\nu},g_{kl}),\;\pi^{ij}+\pi^{\lambda}
    \frac{\partial v_{\lambda}}{\partial g_{ij}}\right\}\nonumber\\
\label{EPS-PB1}
 &=&\frac{\partial V_{\mu}}{\partial g_{0\rho}}\;
    \frac{\partial v_{\lambda}}{\partial g_{ij}}\delta_{\rho}^{\lambda}
  +\frac{\partial V_{\mu}}{\partial g_{kl}}\;\frac12\left(\delta_k^i\delta_l^j+\delta_l^j\delta_k^i\right)
  =\frac{\partial V_{\mu}}{\partial g_{0\lambda}}\;\frac{\partial v_{\lambda}}{\partial g_{ij}}
  +\frac{\partial V_{\mu}}{\partial g_{ij}}=0.
\end{eqnarray}
To give another example, let us check the following bracket:
\begin{equation}
\label{EPS-PB2}
\left.\{N_{\mu},\,\Pi^{\nu}\}\right|_{g_{\lambda\rho},p^{\sigma\tau}}
  =\frac{\partial N_{\mu}}{\partial g_{0\rho}}\;\frac{\partial \Pi^{\nu}}{\partial\pi^{\rho}}
  =\left\{V_{\mu}(g_{0\rho},g_{ij}),\;\pi^{\lambda}
    \frac{\partial v_{\lambda}}{\partial N_{\nu}}\right\}
  =\frac{\partial V_{\mu}}{\partial g_{0\rho}}\;\frac{\partial v_{\rho}}{\partial N_{\nu}}
  =\delta_{\mu}^{\nu}.
\end{equation}
The rest of the brackets can be checked by analogy. This completes the proof of canonicity of the transformation (\ref{ADMl-tr}) for the full gravitational theory.

\section{Discussion}
A starting point for the present investigation was the paper \cite{KK} and the statement made by its authors that components of metric tensor and the ADM variables are not related by a canonical transformation. However, it is misunderstanding to pose the question about canonicity of the transformation (\ref{ADM-tr}) which involves, from the viewpoint of the Dirac approach, non-canonical variables. Let us remind that Dirac himself consider these variables, $g_{0\mu}$ (along with the zero component of vector potential of electromagnetic field $A_0$) as playing the role of Lagrange multipliers while the phase space in his approach includes pairs of generalized coordinates and momenta for which corresponding velocities can be expressed through the momenta.

We should remember also that the Einstein equations were originally formulated in Lagrangian formalism. Dirac's Hamiltonian formulation for gravity is equivalent to Einstein's formulation {\it at the level of equations}. It means that Hamiltonian equations for canonical variables (in Dirac's sense) are equivalent to the ($ij$) Einstein equations, and the gravitational constraints are equivalent to the ($0\mu$) Einstein equations. On the other hand, it implies that a group of transformations in Hamiltonian formalism must involve the full group of gauge transformations of the original theory. However, in the limits of the Dirac approach we fail to  construct a generator that would produce correct transformations for all variables. {\it We inevitably have to modify the Dirac scheme}, and attempts to do it were presented yet in \cite{Cast,BRR}. Therefore, we cannot consider the Dirac approach as fundamental and undoubted.

The ADM formulation of Hamiltonian dynamics for gravity is, first of all, the choice of parametrization, which is preferable because of its geometrical interpretation. There is no any special "ADM procedure": Arnowitt, Deser and Misner constructed the Hamiltonian dynamics following exactly the Dirac scheme, just making use of another variables. The fact that two Hamiltonian formulations (both according to the Dirac scheme, but the one for original variables and the other for the ADM variables) are not related by canonical transformations, should not lead to any bad-grounded conclusions like the one made in \cite{KK}, p. 68, that the gravitational Lagrangian used by Dirac and the ADM Lagrangian are not equivalent. At the Lagrangian level, the transition to the ADM variables is nothing more as a change of variables in the Einstein equations, and there are no mathematical rules that would prohibit such change of variables. It is the Lagrangian formulation of General Relativity which is original and fundamental while its Hamiltonian formulation still remains questionable, in spite of fifty years passed after Dirac's paper \cite{Dirac3}. The extended phase space approach treating all degrees of freedom on an equal footing may be a real alternative to the Dirac generalization of Hamiltonian dynamics.

The example considered in Section 4 shows that the BRST charge can play the role of a sought generator in extended phase space. Nevertheless, the algorithm suggested by BFV for constructing the BRST charge again relies upon the algebra of constraints. Even for the model from Section 4 it would not lead to the correct result (\ref{tilN_trasf}). Another way is to construct the BRST charge as a conserved quantity based on BRST-invariance of the action and making use of the first Noether theorem. This method works satisfactory for simple models with a given symmetry. Below we mentioned that the gravitational Lagrangian density can be modified for the primary constraints to take the simplest form $\pi^{\mu}=0$ without affecting the equation of  motion. However, after this modification the full action may not be BRST-invariant. Some authors (see, for example, \cite{Hall,Hennaux}) use some boundary conditions to exclude total derivatives and ensure BRST-invariance. The boundary conditions (as a rule, these are trivial boundary conditions for ghosts and $\pi^{\mu}$) correspond to asymptotic states and are well-grounded in ordinary quantum field theory. This way does not seem to be general enough, and for gravitational field the justification of the boundary conditions, as well as the control of BRST-invariance of the action, requires special study.

In \cite{Dirac3} Dirac pointed out that ``any dynamical theory must first be put in the Hamiltonian form before one can quantize it''. Based upon Hamiltonian dynamics in extended phase space, a new approach to quantum theory of gravity has been proposed in \cite{SSV1,SSV2}. Ir was argued that it is impossible to construct a mathematically consistent quantum theory of gravity without taking into account the role of gauge degrees of freedom in description of quantum gravitational phenomena from the point of view of different observers. The present paper show that even at the classical level gauge degrees of freedom cannot be excluded from consideration. As we have seen, the extension of phase space by introducing the missing velocities changes the relations between the ``old'' and ``new'' momenta (see (\ref{relat2})). As a consequence, the transformation (\ref{ADMl-tr}) is canonical. In that way, we consider extended phase space not just as an auxiliary construction which enables one to compensate residual degrees of freedom and regularize a path integral, as it was in the Batalin -- Fradkin -- Vilkovisky approach \cite{BFV1,BFV2,BFV3}, but  rather as a structure that ensures equivalence of Hamiltonian dynamics for a constrained system and Lagrangian formulation of the original theory.

\section*{Acknowledgements}
I would like to thank Giovanni Montani and Francesco Cianfrani for attracting my attention to the paper \cite{KK} and discussions.

\end{document}